# Footprint Tracker: reviewing lifelogs and reconstructing daily experiences

**Rúben Gouveia, Evangelos Niforatos, Evangelos Karapanos**
Madeira Interactive Technologies Institute, University of Madeira
Caminho da Penteada, Funchal, 9020-1050, Madeira, Portugal
{ruben.gouveia, evangelos.niforatos, ekarapanos}@m-iti.org

**ABSTRACT**
With the increasing emphasis on how mobile technologies are experienced in everyday life, researchers are increasingly emphasizing the use of in-situ methods such as Experience Sampling and Day Reconstruction. In our line of research we explore the concept of Technology-Assisted Reconstruction, in which passively logged behavior data assist in the later reconstruction of daily experiences. In this paper we introduce *Footprint tracker*, a web application that supports participants in reviewing lifelogs and reconstructing their daily experiences. We focus on three kinds of data: *visual* (as captured through Microsoft's sensecam), *location*, and *context* (i.e., SMS and calls received and made). We describe how Footprint Tracker supports the user in reviewing these lifelogs and outline a field study that attempts to inquire into whether and how this data support reconstruction from memory.

**ACM Classification Keywords**
H5.2. User Interfaces: Evaluation/methodology

## INTRODUCTION
The increasing emphasis on how mobile technologies are experienced in everyday life has resulted in an increased interest in in-situ measurement and, in particular, the Experience Sampling Method (ESM) [1].

ESM is considered as the gold standard of in-situ measurement [2] as it samples experiences and behavior at the moment of their occurrence, thus reducing memory and social biases in self-reporting. However, ESM entails important drawbacks, such as disrupting user's activity and imposing high burden to participants [3].

Motivated by these drawbacks, Daniel Kahneman and colleagues proposed the Day Reconstruction Method (DRM) [2], a retrospective self-report protocol that aims at increasing users' accuracy in reconstructing their experiences at the end of a studied day. It does so by imposing a chronological order in reconstruction, thus providing a temporal context for the recall of each experience. DRM has been found to provide a reasonably good approximation to experience sampling data [2] and the method has been well adopted also in the HCI community (see [4,10] for a review).

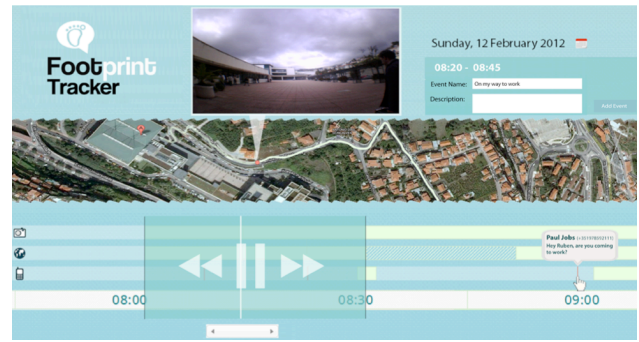

Figure 1. Footprint Tracker enables reviewing visual, location, and context (i.e., SMS and calls made and received) data, describing activities performed throughout the day.

In our line of research we attempt to contribute towards a next step in the field of momentary assessment, that of *technology-assisted reconstruction* (TAR) [4]. TAR consists of passively logging users' behaviors throughout the day with mobile sensor technology and employing these data to assist the reconstruction of one's daily activities and experiences.

In this paper we present Footprint Tracker, a web 2.0 application that assists individuals in reviewing logged behavior data. We describe how Footprint Tracker was motivated by recent research in life-logging and describe a field study that will attempt to inquire into how visual, location and context data support the reconstruction of daily experiences.

## LIFELOGS AND EXPERIENCE RECONSTRUCTION
Life-logging, a vision for a future in which "technology will enable a total recall of our lives through total capture of personally relevant information" [5], has attracted substantial interest over the last decade. A wealth of systems have been proposed tapping upon a variety of lifelogs from passive *visual* data (notably with Microsoft's Sensecam), *location* and other *context* data (such as mobile activity – e.g, SMS and calls), *biometric* data, such as heart rate, galvanic skin response and others.

Yet, while various systems have been proposed, often these have relied on speculations on how this data supports reconstruction from memory, exposing a lack of empirical evidence on how lifelogs mediate memory [5,6].

Our work on Technology-Assisted Reconstruction [4] and Footprint Tracker in particular is motivated by a recent theory of how individuals recall emotions experienced in past events. This theory assumes that the "emotional experience can neither be stored nor retrieved" [8, p. 935]. Instead, it assumes that people first retrieve contextual details from episodic memory and then infer emotions on the basis of this information (e.g., I recall myself screaming and having my hands raised while on the rollercoaster, thus I infer an experience of high arousal). It is thus suggested that through increasing the amount of recalled contextual cues from episodic memory, one could increase individuals' accuracy in recalling the exact emotions experienced during the event (see [9] for a more elaborate discussion on the topic).

## FOOTPRINT TRACKER

Motivated by Robinson and Clore's [8] theory, Footprint tracker employs lifelogs in attempting to assist individuals to recall information from episodic memory. This is then assumed to increase their ability to infer the emotions experienced in a particular event with greater accuracy. Footprint Tracker places substantial emphasis on the temporal representation of each activity and attempts to impose a chronological order in reconstruction. Karapanos et al. [9] showed that even with subtle variations in the reconstruction process (e.g., imposing a chronological order) one may affect individuals' ability to reconstruct experiences that lie 6-12 months in the past. Footprint tracker currently supports three kinds of lifelogs:

a) *visual* data as captured from the Vicon Revue camera (a.k.a. sensecam) have been found to cue directly information from episodic memory [6,7],

b) *location* data representing both significant location (e.g., ones that an individual has spent more than 5 minutes in a 50 meter radius), as well as transitions. Location data have been previously found to mediate memory through tapping to daily routines [7],

c) *context* data reflected in SMS and calls made and received throughout the day, which are assumed to cue the recall of particular moments in time.

The interface is split down into two main sections:

a) *timeline pane* (bottom) highlighting the presence of visual data (green indicates presence of visual data, and blue an absence of data), location data (solid green indicates no transitions, dashed green represents transition and blue indicates an absence of location data), and context data (solid green represents a phone call and its respective duration, a red line targets incoming messages, while blue means lack of cell phone data). These four bars are interactive, allowing users to choose (and adjust) a period of time where they wish to view data (see fig 1),

b) *data pane* (top) that depicts location and visual data. By selecting a period of time in the *timeline pane*, images and respective GPS location will be loaded in the *visual data pane*. Users can navigate through these using the pause/play, fast-forward/rewind controls. Also, significant locations are represented as circles. The bigger the circle, the more time spent at a certain location.

## CONCLUSION AND ONGOING WORK

Founded upon our vision of Technology-Assisted Reconstruction, Footprint tracker aims at supporting the reconstruction of daily experiences and whereabouts through the revue of life-logging data. Footprint Tracker is distinct than many lifelogging approaches since it is aimed as a methodological tool for the in-situ evaluation of ubiquitous computing applications, and for shorter periods of time than what is common in life-logging applications. Our ongoing study aims at inquiring into whether and how visual, location and context data support individuals in recalling their daily experiences with greater accuracy. Experience Sampling is used as ground truth and the effectiveness of Footprint Tracker will be compared against the current state-of-the-art in retrospective evaluation, the Day Reconstruction Method.